\documentclass[fleqn,usenatbib]{mnras}

\usepackage[table]{xcolor}
\usepackage[T1]{fontenc}
\usepackage{ae,aecompl}
\usepackage{xtab}
\usepackage{courier}
\usepackage{multirow}
\usepackage{times}
\usepackage{upgreek}
\usepackage{graphicx}
\usepackage{epstopdf}
\usepackage{amssymb}
\usepackage{amsmath}
\usepackage[section]{placeins}

\usepackage{newtxtext,newtxmath}
\usepackage[T1]{fontenc}
\usepackage{ae,aecompl}

\usepackage{tabulary}
\usepackage{graphicx}	% Including figure files
\usepackage{amsmath}	% Advanced maths commands
\usepackage{amssymb}	% Extra maths symbols

\usepackage{soul}
\newcommand{\uHz}{$\upmu$Hz}

 \newcommand{\cmss}{\,\mbox{$\mbox{cm}\,\mbox{s}^{-2}$}}    % centimetres/second^2
\newcommand{\tess}{{\em TESS }}

\newcommand{\gaia}{{\em Gaia}}

\newcommand{\teff}{T$_{\rm eff}$}
\newcommand{\loggcms}{$\log{(g/\cmss)}$}

\definecolor{light1}{gray}{0.75}
\definecolor{light2}{gray}{0.85}
\definecolor{light3}{gray}{0.95}
\title[\tess\ sdBVs in both ecliptic hemispheres]{A search for variable subdwarf B stars in \tess\ Full Frame Images\\
III. An update on variable targets in both ecliptic hemispheres -- contamination analysis and new sdB pulsators}
\author[S.K.\,Sahoo et al.]{S.\,K.\,Sahoo$^{1,2}$\thanks{E-mail: sumanta.kumar27@gmail.com},
A.\,S.\,Baran$^{2,3,4}$,
H.L.\,Worters$^{5}$,
P.\,N\'emeth$^{2,6,7}$
and
D.\,Kilkenny$^{8}$
\\
$^{1}$Nicolaus Copernicus Astronomical Centre of the Polish Academy of Sciences, ul. Bartycka 18, 00-716 Warsaw, Poland\\
$^{2}$ARDASTELLA Research Group\\
$^{3}$Astronomical Observatory, University of Warsaw, Al. Ujazdowskie 4, 00-478 Warszawa, Poland\\
$^{4}$Department of Physics, Astronomy, and Materials Science, Missouri State University, Springfield, MO\,65897, USA\\
$^{5}$South African Astronomical Observatory, Observatory 7935, South Africa\\
$^{6}$Astronomical Institute of the Czech Academy of Sciences, Fri\v{c}ova 298, CZ-251\,65 Ond\v{r}ejov, Czech Republic\\
$^{7}$Astroserver.org, F\H{o} t\'er 1, 8533 Malomsok, Hungary\\
$^{8}$Department of Physics and Astronomy, University of the Western Cape, Private Bag X17, Bellville 7535, South Africa\\
}
\date{Accepted XXX. Received YYY; in original form ZZZ}

\pubyear{2020}

\begin{document}
\label{firstpage}
\pagerange{\pageref{firstpage}--\pageref{lastpage}}
\maketitle

\begin{abstract}
We present an update on the variable star survey performed on the \tess\ 30\,min Full Frame Image (FFI) data reported by our first two papers in this series. This update includes a contamination analysis in order to identify false positives and analysis of the \tess\ 10\,min FFI data collected during Years\,3 and 4 of the mission. We clarify the variability status of 2\,995 targets identifying 1\,403 variable stars. In addition, we spectroscopically classify 24 pre-filtered targets sampled with the 10\,min FFI data and discover 11 new sdB pulsators. Future follow-up space- and/or ground-based data of variables reported here, to identify the nature of their variability and reveal spectroscopic parameters of the stars, would complement this work.
\end{abstract}


%%%%%%%%%%%%%%%%%%%%%%%%%%%%%%%%%%%%%%%%%%%%%%%%%%
%%%%%%%%%%%%%%%%% BODY OF PAPER %%%%%%%%%%%%%%%%%%

\section{Introduction}
\citet{sahoo20b} (Paper\,I) and \citet{baran21a} (Paper\,II) presented their results of variability checks of the most promising subdwarf B (sdB) candidates found in the \citet{geier19} and \citet{geier20} catalogs. The former authors pre-selected 45\,674 targets and used the Full Frame Images (FFI) collected by the \tess\ mission in the Southern Ecliptic Hemisphere (SEH) during Year\,1 and in the Northern Ecliptic Hemisphere (NEH) during Year\,2. As a result, 2\,313 new variable targets in both ecliptic hemispheres were listed.

It is a well known feature of the \tess\ CCDs that an individual pixel has a 21\,arcsec square projection on the sky. This makes contamination a serious problem. A contaminating star contributes by either increasing the average flux level in an aperture, thus lowering the apparent amplitude of flux variation in the target, or -- if variable -- makes a constant flux target star a false positive variable. The former issue can be corrected by removing the estimated amount of a contaminating star's flux and this is fairly well achieved in the Pre-search Data Conditioning fluxes. The latter issue can only be resolved by checking the variability of all contaminators around the target of interest. We also found this method useful to resolve variability in case both the target of interest and contaminator are variables.

In this work, we aimed to remove false positives from our list of 2\,313 new variable objects. We searched all pixels within the target masks of all 2\,313 objects listed in Papers I and II. In addition, for the most promising pulsating subdwarf candidates, we retrieved the new light curves from the FFI collected during Years\,3 and 4, which are sampled at a 10\,min cadence. This cadence extends the Nyquist frequency up to 833.3\,\uHz, which allows us to cover the entire gravity mode region in an amplitude spectrum\footnote{In hot subdwarfs, pressure or p-mode pulsations typically have periods of a few minutes; gravity or g-modes have periods of the order of an hour. See, for example, \citet{heber16}.}. By promising candidates, we mean objects that in Papers I and II show signals close to the Nyquist frequency of 277.7\,\uHz \, resulting from the 30\,min cadence of observation used during Years\,1 and 2.

To summarize the findings reported in Paper I and II, there are 15 pulsating subdwarf B star candidates, 79 variable (other than pulsating) sdB stars, 33 variable subdwarf candidates and 2\,186 other variable stars, including 123 stars with non-sd classification.

Data are downloaded and processed in the same way as explained in Papers I and II. For a convenient comparison, the Tables presented here will preserve the listing order from the previous papers of this series, however some of the tables from Paper\,I are merged to be consistent with those in Paper\,II. They also contain the most important information from Papers I and II with additional information resulting from this work. In the following sections, we explain the details of the contamination analysis and provide the results of this work.

\section{contamination analysis}
\label{sec:contamination}
We defined each target mask as a square of 11\,$\times$\,11\,pixels, which provides enough pixels to clarify contamination, if any. We used our custom \texttt{PYTHON} scripts to retrieve fluxes separately from each pixel in a 3\,$\times$\,3\,pixel square centered at a given target. Then, for each of the nine pixels, we calculated an amplitude spectrum and checked if the signals reported in Papers\,I and II is detected in pixels that overlap with the location of the target. We employed either PanSTARRS or DSS images and overlapped them with target pixel files (TPF) created from the \texttt{TESSCUT} tool \citep{brasseur19}. In specific cases, that is, bright stars in target masks, even though far enough from our targets not to contaminate them, we calculated the amplitude spectra in pixels covering these stars to check for their variability. As a result of our contamination analysis, we derived a few common cases, listed below (the columns refer to Tables in the online materials):

\begin{table*}
\centering
\caption{Southern Ecliptic Hemisphere (SEH)\,--\,28 objects classified as sdBs. Only the first five objects are listed while the full table can be found in the online materials.}
\label{tab:tab1}
\rowcolors{1}{}{light3}
{\begin{tabular}{clrlccl}
\hline
\rowcolor{light1}
& & & & \multicolumn{1}{c}{G} & \multicolumn{1}{c}{Period} & \multicolumn{1}{c}{Variable} \\
\rowcolor{light1}
\multicolumn{1}{c}{\multirow{-2}{*}{No.}} & \multicolumn{1}{c}{\multirow{-2}{*}{\gaia\ DR2}} & \multicolumn{1}{c}{\multirow{-2}{*}{TIC}} & \multicolumn{1}{c}{\multirow{-2}{*}{Name}} & \multicolumn{1}{c}{[mag]} & \multicolumn{1}{c}{[d]} & \multicolumn{1}{c}{contaminator}\\
\hline
\rowcolor{light2}
\multicolumn{7}{c}{sdBVs}\\
\hline
1 & 3129751228471383808 & 237597052 & TYC 161-49-1 & 11.14 & 0.05-0.1 & none\\
2 & 3159937564294110080 & 262753627 & TYC 770-941-1 & 12.46 & 0.04-0.08 & none\\
\hline
\rowcolor{light2}
\multicolumn{7}{c}{Phased lightcurves}\\
\hline
1 & 2333936291513550336 & 12379252 & Ton S138 & 16.01 & 0.2648 & none\\
2 & 2385348183917624448 & 9035375 & PHL 460 & 12.21 & 0.4734 & none\\
3 & 2969438206889996160 & 139397815 & - & 13.61 & 0.2746 & none\\
\hline
\end{tabular}}
\end{table*}

\begin{table*}
\centering
\caption{Northern Ecliptic Hemisphere (NEH)\,--\,53 objects classified as sdBs. Only the first five objects are listed while the full table can be found in the online materials.}
\label{tab:tab2}
\rowcolors{1}{}{light3}
{\begin{tabular}{clrlccl}
\hline
\rowcolor{light1}
& & & & \multicolumn{1}{c}{G} & \multicolumn{1}{c}{Period} & \multicolumn{1}{c}{Variable} \\
\rowcolor{light1}
\multicolumn{1}{c}{\multirow{-2}{*}{No.}} & \multicolumn{1}{c}{\multirow{-2}{*}{\gaia\ DR2}} & \multicolumn{1}{c}{\multirow{-2}{*}{TIC}} & \multicolumn{1}{c}{\multirow{-2}{*}{Name}} & \multicolumn{1}{c}{[mag]} & \multicolumn{1}{c}{[d]} & \multicolumn{1}{c}{contaminator}\\
\hline
\rowcolor{light2}
\multicolumn{7}{c}{sdBVs}\\
\hline
1 & 1028374599849118976 & 802232206 & SDSSJ082428.41+512601.6 & 18.72 & 0.1734 & CRTS J082433.5+512441 (EB)\\
2 & 127674641678296704 & 353892824 & KUV02281+2730 & 15.15 & 0.0487 & none\\
3 & 1345049483546987904 & 159850392 & GALEXJ17566+4125 & 14.28 & 0.08 - 0.13 & none\\
4 & 1469357759922416256 & 321423000 & SDSSJ132432.37+320420.9 & 16.64 & 0.0907 & TIC 321422994\\
5 & 1495329392800826624 & 23746001 & PG1350+372 & 14.31 & 0.06 - 0.09 & none\\
\hline
\end{tabular}}
\end{table*}

\begin{table*}
\centering
\caption{NEH\,--\,33 objects classified as sds. Only the first five objects are listed while the full table can be found in the online materials.}
\label{tab:tab3}
\rowcolors{1}{}{light3}
{\begin{tabular}{clrlccl}
\hline
\rowcolor{light1}
& & & & \multicolumn{1}{c}{G} & \multicolumn{1}{c}{Period} & \multicolumn{1}{c}{Variable} \\
\rowcolor{light1}
\multicolumn{1}{c}{\multirow{-2}{*}{No.}} & \multicolumn{1}{c}{\multirow{-2}{*}{\gaia\ DR2}} & \multicolumn{1}{c}{\multirow{-2}{*}{TIC}} & \multicolumn{1}{c}{\multirow{-2}{*}{Name}} & \multicolumn{1}{c}{[mag]} & \multicolumn{1}{c}{[d]} & \multicolumn{1}{c}{contaminator}\\
\hline
\rowcolor{light2}
\multicolumn{7}{c}{sdVs}\\
\hline
1 & 1906485375099435136 & 259091223 & FBS2209+354 & 14.30 & 0.07-0.13 & no signal in individual pixels\\
2 & 1952553606634620928 & 407657360 & LAMOSTJ214600.31+372119.7 & 14.66 & 0.045 - 0.1 & TIC 407657373\\
3 & 2041883531914920064 & 20688004 & GALEXJ18578+3048 & 13.73 & 0.1726 & ATO J284.4865+30.8044 (EB candidate)\\
4 & 2128012018629286144 & 1882679963 & KeplerJ19352+4555 & 17.16 & 0.1766 & ATO J293.8168+45.8972 (EB candidate)\\
5 & 237650985848157312 & 194781979 & LAMOSTJ032717.71+410344.5 & 10.19 & 0.1594 & none\\
\hline
\end{tabular}}
\end{table*}

\begin{table*}
\centering
\caption{SEH\,--\,83 pulsator candidates. Only the first five objects are listed while the full table can be found in the online materials.}
\label{tab:tab4}
\rowcolors{1}{}{light3}
{\begin{tabular}{clrcl}
\hline
\rowcolor{light1}
& & & \multicolumn{1}{c}{G} & \multicolumn{1}{c}{Variable}\\
\rowcolor{light1}
\multicolumn{1}{c}{\multirow{-2}{*}{No.}} & \multicolumn{1}{c}{\multirow{-2}{*}{\gaia\ DR2}} & \multicolumn{1}{c}{\multirow{-2}{*}{TIC}} & \multicolumn{1}{c}{[mag]} & \multicolumn{1}{c}{contaminator}\\
\hline
1 & 2921500461998485248 & 744231977 & 18.31 & TYC 6526-2198-1\\
2 & 2927637764107094272 & 744958933 & 18.67 & ATO J106.3086-23.5750\\
3 & 3062196993541803904 & 754827446 & 17.45 & TYC 4817-751-1\\
4 & 3087146252404755584 & 257068255 & 15.08 & none\\
5 & 3111790534231122944 & 284329074 & 15.26 & TYC 163-370-1\\
\hline
\end{tabular}}
\end{table*}

\begin{table*}
\centering
\caption{NEH\,--\,30 pulsator candidates. Only the first five objects are listed while the full table can be found in the online materials.}
\label{tab:tab5}
\rowcolors{1}{}{light3}
{\begin{tabular}{clrcl}
\hline
\rowcolor{light1}
& & & \multicolumn{1}{c}{G} & \multicolumn{1}{c}{Variable}\\
\rowcolor{light1}
\multicolumn{1}{c}{\multirow{-2}{*}{No.}} & \multicolumn{1}{c}{\multirow{-2}{*}{\gaia\ DR2}} & \multicolumn{1}{c}{\multirow{-2}{*}{TIC}} & \multicolumn{1}{c}{[mag]} & \multicolumn{1}{c}{contaminator}\\
\hline
1 & 1422182595056481536 & 320525680 & 13.87 & no signal in individual pixels\\
2 & 1943952161530528256 & 431548978 & 15.61 & no signal in individual pixels\\
& & & & 34.14 \uHz: none; \\
\rowcolor{white}
\rowcolors{25}{}{light3}
\multirow{-2}{*}{3} & \multirow{-2}{*}{1974973679520560896} & \multirow{-2}{*}{311792028} & \multirow{-2}{*}{15.88} & 95.37 \uHz+harmonics: TYC 3605-1317-1 (V) \\
\rowcolor{light3}
4 & 1988552407605096320 & 66784300 & 14.88 & TIC 66784249\\
5 & 1991879937806406656 & 2044241813 & 18.47 & NSVS 1502401 (EB)\\
\hline
\end{tabular}}
\end{table*}

\begin{table*}
\centering
\caption{SEH\,--\,83 eclipsing binaries. Only the first five objects are listed while the full table can be found in the online materials.}
\label{tab:tab6}
\rowcolors{1}{}{light3}
{\begin{tabular}{clrccl}
\hline
\rowcolor{light1}
& & & \multicolumn{1}{c}{G} & \multicolumn{1}{c}{Period} & \multicolumn{1}{c}{Variable} \\
\rowcolor{light1}
\multicolumn{1}{c}{\multirow{-2}{*}{No.}} & \multicolumn{1}{c}{\multirow{-2}{*}{\gaia\ DR2}} & \multicolumn{1}{c}{\multirow{-2}{*}{TIC}} & \multicolumn{1}{c}{[mag]} & \multicolumn{1}{c}{[d]} & \multicolumn{1}{c}{contaminator}\\
\hline
\rowcolor{light2}
\multicolumn{6}{c}{32 eclipsing binaries that show both primary and secondary eclipses}\\
\hline
1 & 2938186341221700480 & 60523137 & 16.23 & 1.2532 & none\\
2 & 3056677303432024960 & 753916356 & 17.97 & 2.4282 & TIC 68060528\\
3 & 4037952609036313728 & 1556986400 & 18.86 & 4.2844 & TIC 368875977\\
4 & 4038037855601783296 & 1557298522 & 17.18 & 2.9735 & TYC 7404-5579-1\\
5 & 4044609357370901632 & 1569961982 & 16.46 & 2.415 & RS Sgr (B3/4IV/V, EB)\\
\hline
\end{tabular}}
\end{table*}

\begin{table*}
\centering
\caption{NEH\,--\,23 eclipsing binaries. Only the first five objects are listed while the full table can be found in the online materials.}
\label{tab:tab7}
\rowcolors{1}{}{light3}
{\begin{tabular}{clrccl}
\hline
\rowcolor{light1}
& & & \multicolumn{1}{c}{G} & \multicolumn{1}{c}{Period} & \multicolumn{1}{c}{Variable} \\
\rowcolor{light1}
\multicolumn{1}{c}{\multirow{-2}{*}{No.}} & \multicolumn{1}{c}{\multirow{-2}{*}{\gaia\ DR2}} & \multicolumn{1}{c}{\multirow{-2}{*}{TIC}} & \multicolumn{1}{c}{[mag]} & \multicolumn{1}{c}{[d]} & \multicolumn{1}{c}{contaminator}\\
\hline
\rowcolor{light2}
\multicolumn{6}{c}{Eclipsing binaries that show only primary eclipses}\\
\hline
& & & & & 49.77 \uHz+harmonics: none; \\
\rowcolor{light3}
\rowcolors{25}{}{white}
\multirow{-2}{*}{1} & \multirow{-2}{*}{1131845039229607680} & \multirow{-2}{*}{459182998} & \multirow{-2}{*}{16.16} & 
\multirow{-2}{*}{0.2344} & 
6.94 \uHz+harmonic: TIC 459183003 \\
\rowcolor{white}
2 & 1417117518648285056 & 1400704733 & 17.03 & 0.3637 & none\\
3 & 1816806183083980288 & 1943324398 & 17.22 & 1.3135 & HD 195052 (F8)\\
4 & 1840900601716813440 & 1951174238 & 18.92 & 1.0329 & TIC 126684646\\
5 & 1846629538332584960 & 15040115 & 11.83 & 0.8099 & none\\
\hline
\end{tabular}}
\end{table*}

\begin{table*}
\centering
\caption{SEH\,--\,273 spectroscopically unclassified variables with phased light curves. Only the first five objects are listed while the full table can be found in the online materials.}
\label{tab:tab8}
\rowcolors{1}{}{light3}
{\begin{tabular}{clrccl}
\hline
\rowcolor{light1}
& & & \multicolumn{1}{c}{G} & \multicolumn{1}{c}{Period} & \multicolumn{1}{c}{Variable} \\
\rowcolor{light1}
\multicolumn{1}{c}{\multirow{-2}{*}{No.}} & \multicolumn{1}{c}{\multirow{-2}{*}{\gaia\ DR2}} & \multicolumn{1}{c}{\multirow{-2}{*}{TIC}} & \multicolumn{1}{c}{[mag]} & \multicolumn{1}{c}{[d]} & \multicolumn{1}{c}{contaminator}\\
\hline
\rowcolor{light2}
\multicolumn{6}{c}{One symmetric maximum}\\
\hline
1 & 2896588449084891136 & 49547169 & 13.28 & 0.3086 & IS CMa (F3V)\\
2 & 2905822663130146688 & 31353391 & 14.03 & 0.8929 & none\\
3 & 2909497952544966272 & 37118148 & 14.28 & 0.2681 & none\\
4 & 2911497105202950400 & 37004041 & 15.16 & 0.2833 & none\\
5 & 2921050693020996864 & 63113578 & 11.45 & 0.4854 & none\\
\hline
\end{tabular}}
\end{table*}

\begin{table*}
\centering
\caption{NEH\,--\,93 spectroscopically unclassified variables with phased light curves. Only the first five objects are listed while the full table can be found in the online materials.}
\label{tab:tab9}
\rowcolors{1}{}{light3}
{\begin{tabular}{clrccl}
\hline
\rowcolor{light1}
& & & \multicolumn{1}{c}{G} & \multicolumn{1}{c}{Period} & \multicolumn{1}{c}{Variable} \\
\rowcolor{light1}
\multicolumn{1}{c}{\multirow{-2}{*}{No.}} & \multicolumn{1}{c}{\multirow{-2}{*}{\gaia\ DR2}} & \multicolumn{1}{c}{\multirow{-2}{*}{TIC}} & \multicolumn{1}{c}{[mag]} & \multicolumn{1}{c}{[d]} & \multicolumn{1}{c}{contaminator}\\
\hline
\rowcolor{light2}
\multicolumn{6}{c}{One symmetric maximum}\\
\hline
1 & 1000519267329142144 & 444946935 & 15.92 & 1.6725 & none\\
2 & 1086341235118052096 & 85158691 & 12.85 & 0.3546 & none\\
3 & 1099487030500185344 & 743328948 & 16.90 & 0.3297 & none\\
4 & 1133795950814826240 & 841399917 & 17.31 & 1.5134 & none\\
5 & 1141625057721183616 & 138400883 & 16.24 & 0.0680 & none\\
\hline
\end{tabular}}
\end{table*}

\begin{table*}
\centering
\caption{SEH\,--\,two novae. This table is also included in the online materials.}
\label{tab:tab10}
\rowcolors{1}{}{light3}
{\begin{tabular}{clrlcl}
\hline
\rowcolor{light1}
& & & & \multicolumn{1}{c}{G} & \multicolumn{1}{c}{Variable} \\
\rowcolor{light1}
\multicolumn{1}{c}{\multirow{-2}{*}{No.}} & \multicolumn{1}{c}{\multirow{-2}{*}{\gaia\ DR2}} & \multicolumn{1}{c}{\multirow{-2}{*}{TIC}} & \multicolumn{1}{c}{\multirow{-2}{*}{Name}} & \multicolumn{1}{c}{[mag]} & \multicolumn{1}{c}{contaminator}\\
\hline
1 & 5207384891323130368 & 735128403 & AH Men & 13.51 & none\\
2 & 6544371342567818496 & 121422158 & RZ Gru & 12.63 & none\\
\hline
\end{tabular}}
\end{table*}

\begin{table*}
\centering
\caption{SEH\,--\,1262 spectroscopically unclassified variables with amplitude spectra. Only the first five objects are listed while the full table can be found in the online materials.}
\label{tab:tab11}
\rowcolors{1}{}{light3}
{\begin{tabular}{clrcl}
\hline
\rowcolor{light1}
& & & \multicolumn{1}{c}{G} & \multicolumn{1}{c}{Variable}\\
\rowcolor{light1}
\multicolumn{1}{c}{\multirow{-2}{*}{No.}} & \multicolumn{1}{c}{\multirow{-2}{*}{\gaia\ DR2}} & \multicolumn{1}{c}{\multirow{-2}{*}{TIC}} & \multicolumn{1}{c}{[mag]} & \multicolumn{1}{c}{contaminator}\\
\hline
1 & 2326333512204996992 & 380826878 & 15.69 & none\\
2 & 2342907791000463232 & 610076106 & 17.18 & [SHM2017] J013.19449-26.56892 (RR Lyr)\\
3 & 2342907962798690944 & 610077229 & 16.68 & [SHM2017] J013.19449-26.56892 (RR Lyr)\\
4 & 2409630520260038784 & 2052262357 & 18.09 & Cl* NGC 7492 C 1306 (RR Lyr)\\
5 & 2410677839445234944 & 111183765 & 14.48 & none\\
\hline
\end{tabular}}
\end{table*}

\begin{table*}
\centering
\caption{NEH\,--\,228 spectroscopically unclassified variables with amplitude spectra. Only the first five objects are listed while the full table can be found in the online materials.}
\label{tab:tab12}
\rowcolors{1}{}{light3}
{\begin{tabular}{clrcl}
\hline
\rowcolor{light1}
& & & \multicolumn{1}{c}{G} & \multicolumn{1}{c}{Variable}\\
\rowcolor{light1}
\multicolumn{1}{c}{\multirow{-2}{*}{No.}} & \multicolumn{1}{c}{\multirow{-2}{*}{\gaia\ DR2}} & \multicolumn{1}{c}{\multirow{-2}{*}{TIC}} & \multicolumn{1}{c}{[mag]} & \multicolumn{1}{c}{contaminator}\\
\hline
1 & 1082306439760979840 & 743148169 & 18.81 & TIC 284473271\\
2 & 1107705772542003200 & 705157619 & 18.09 & no signal in individual pixels\\
3 & 1108642968765677696 & 743476657 & 18.74 & V486 Cam (RR Lyr)\\
4 & 1112770367915047424 & 705166070 & 17.41 & none\\
5 & 1113516073020001152 & 705175423 & 18.32 & TIC 468921975\\
\hline
\end{tabular}}
\end{table*}

\begin{table*}
\centering
\caption{SEH\,--\,76 non-sdB classified variables. Only the first five objects are listed while the full table can be found in the online materials.}
\label{tab:tab13}
\rowcolors{1}{}{light3}
{\begin{tabular}{clrcccl}
\hline
\rowcolor{light1}
& & & \multicolumn{1}{c}{G} & & \multicolumn{1}{c}{Period} & \multicolumn{1}{c}{Variable} \\
\rowcolor{light1}
\multicolumn{1}{c}{\multirow{-2}{*}{No.}} & \multicolumn{1}{c}{\multirow{-2}{*}{\gaia\ DR2}} & \multicolumn{1}{c}{\multirow{-2}{*}{TIC}} & \multicolumn{1}{c}{[mag]} & \multicolumn{1}{c}{\multirow{-2}{*}{spT}} & \multicolumn{1}{c}{[d]} & \multicolumn{1}{c}{contaminator}\\
\hline
\rowcolor{light2}
\multicolumn{7}{c}{Pulsators}\\
\hline
1 & 2969201399574096128 & 708596809 & 11.30 & A0IV/V & - & no signal in individual pixels\\
2 & 3109409266919646976 & 168595004 & 10.70 & A5 & - & none\\
3 & 3115125211261708032 & 293137161 & 11.01 & A0 & - & none\\
4 & 3344114626761364224 & 437889214 & 10.02 & B5 & - & none\\
5 & 3396397877830881792 & 247513086 & 8.17 & A0 & - & none\\
\hline
\end{tabular}}
\end{table*}

\begin{table*}
\centering
\caption{NEH\,--\,46 non-sdB classified variables. Only the first five objects are listed while the full table can be found in the online materials.}
\label{tab:tab14}
\rowcolors{1}{}{light3}
{\begin{tabular}{clrcccl}
\hline
\rowcolor{light1}
& & & \multicolumn{1}{c}{G} & & \multicolumn{1}{c}{Period} & \multicolumn{1}{c}{Variable} \\
\rowcolor{light1}
\multicolumn{1}{c}{\multirow{-2}{*}{No.}} & \multicolumn{1}{c}{\multirow{-2}{*}{\gaia\ DR2}} & \multicolumn{1}{c}{\multirow{-2}{*}{TIC}} & \multicolumn{1}{c}{[mag]} & \multicolumn{1}{c}{\multirow{-2}{*}{spT}} & \multicolumn{1}{c}{[d]} & \multicolumn{1}{c}{contaminator}\\
\hline
\rowcolor{light2}
\multicolumn{7}{c}{Pulsators}\\
\hline
1 & 1625627602365544832 & 198209459 & 13.53 & B2 & - & none\\
2 & 1713032695100155904 & 298091568 & 15.00 & B4.1 & - & none\\
3 & 1861191062326013696 & 1955410399 & 10.31 & A0 & - & none\\
& & & & & & 11 \uHz+harmonics: none; \\
\rowcolor{white}
\rowcolors{25}{}{light3}
\multirow{-2}{*}{4} & \multirow{-2}{*}{2077737678383889408} & \multirow{-2}{*}{270610177} & \multirow{-2}{*}{11.77} & 
\multirow{-2}{*}{Be} & 
\multirow{-2}{*}{-} & 
46 - 70 \uHz: UCAC4 663-077912 (PulV) \\
\rowcolor{light3}
5 & 2132171608553758336 & 279919275 & 13.46 & B3V & - & none\\
\hline
\end{tabular}}
\end{table*}

\begin{figure*}
\includegraphics[width=0.68\textwidth]{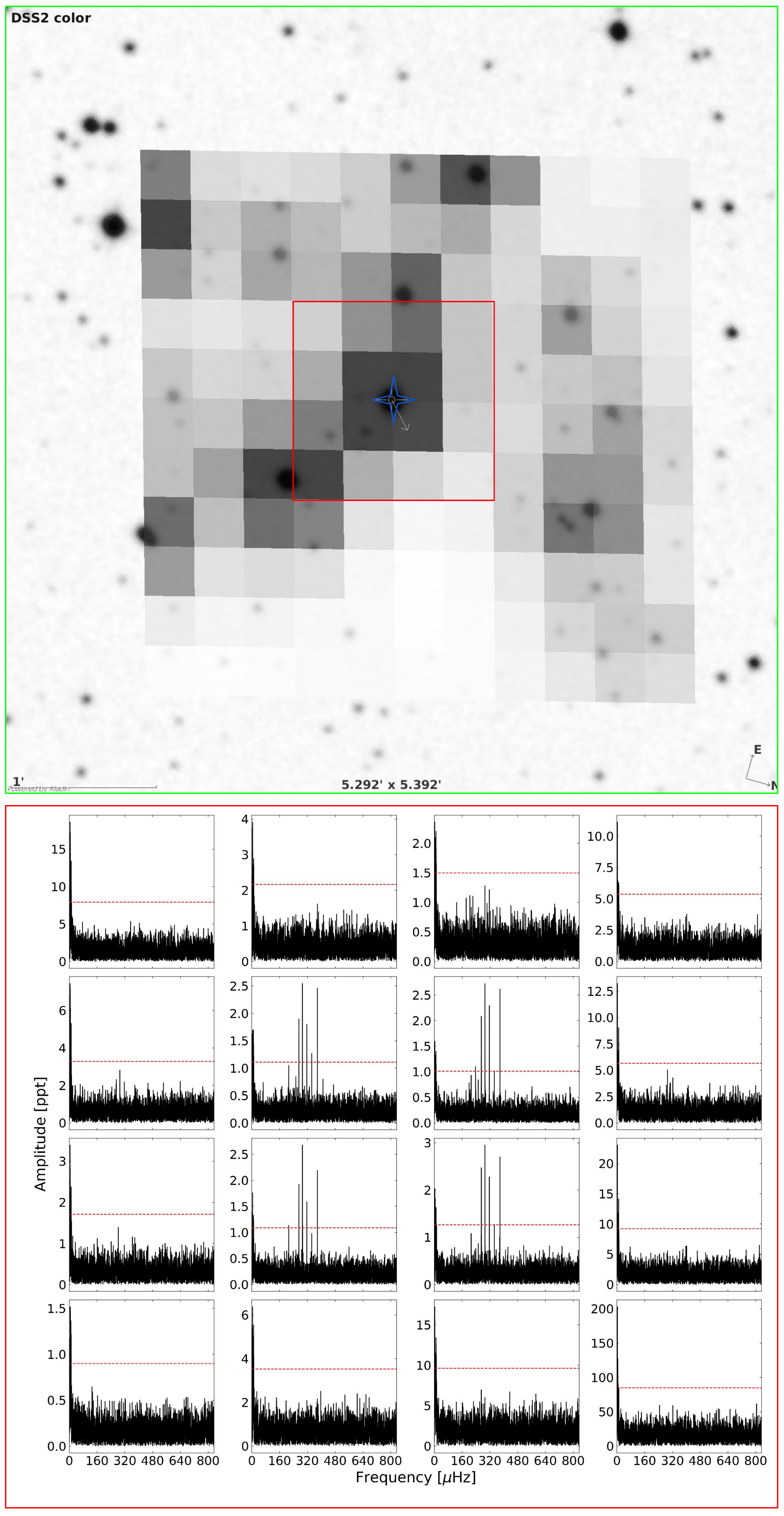}
\caption{Contamination analysis for TIC\,323174439, representing an uncontaminated object. Top panel: target mask overlapped with a DSS2 image. The blue star indicates the target. The red box defines the pixels for which the amplitude spectra were calculated. Bottom panel: Amplitude spectra of the 16 pixels within the red box in the top panel.}
\label{fig:323174439}
\end{figure*}

\begin{figure*}
\includegraphics[width=0.68\textwidth]{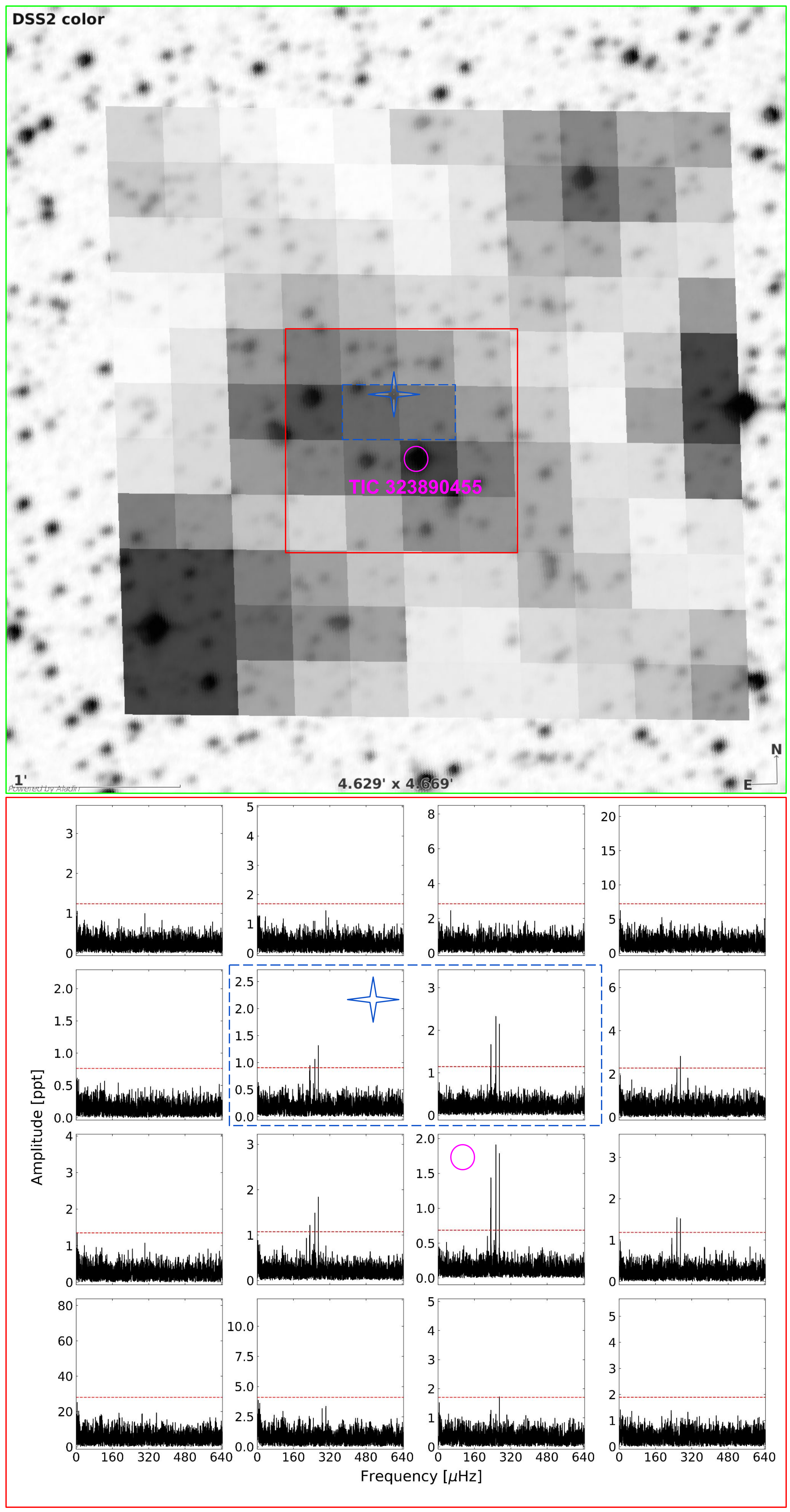}
\caption{Contamination analysis for TIC\,1514267365, representing a false positive case. The marks and colors are the same as in Figure\,\ref{fig:323174439}. The blue rectangle represents the aperture used in Paper\,I. The dominant signal comes from pixels centered on TIC\,323890455 (magenta circle in the top panel) and overlaps with the aperture used in Paper\,I.}
\label{fig:1514267365}
\end{figure*}

\begin{figure*}
\includegraphics[width=0.73\textwidth]{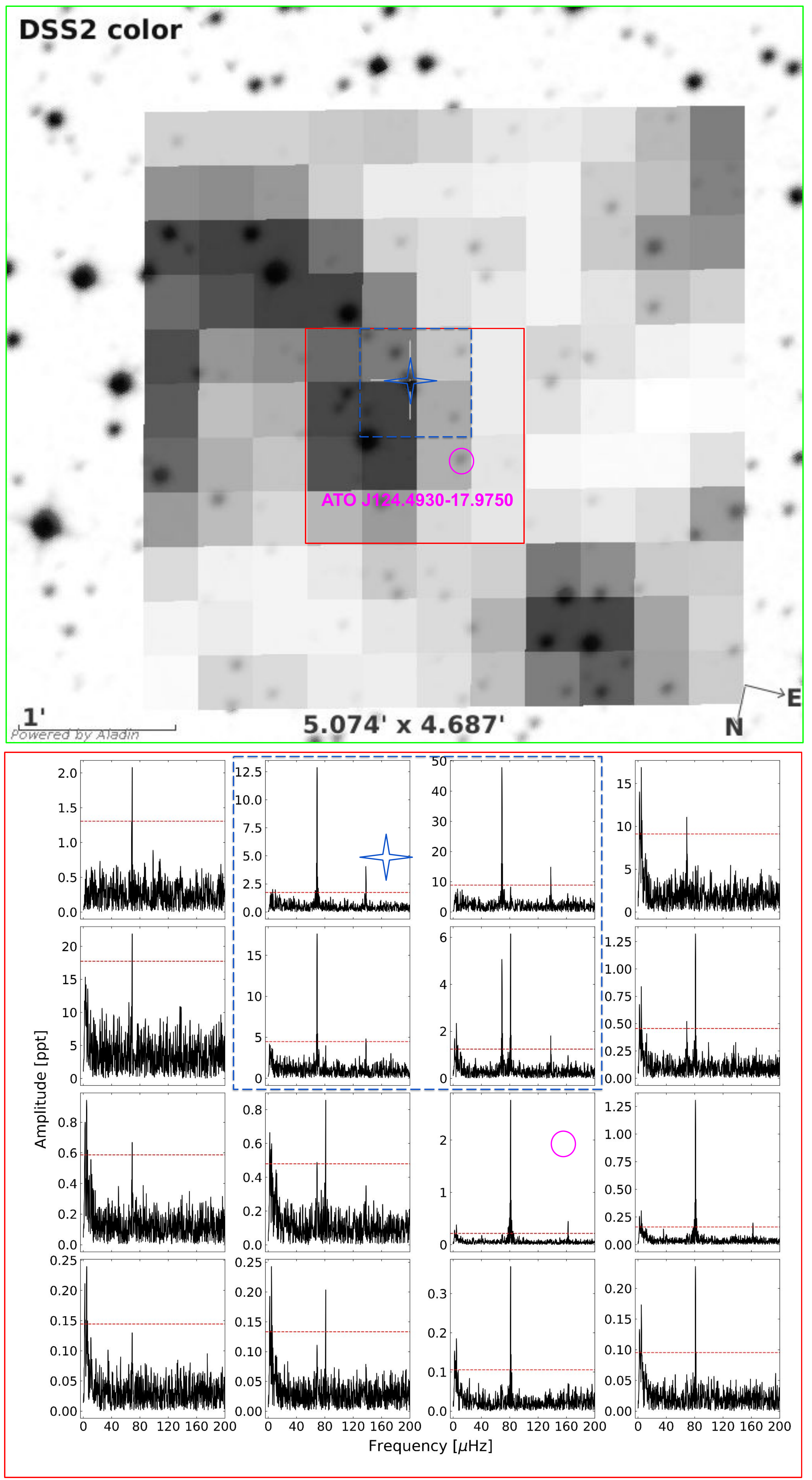}
\caption{Contamination analysis for TIC\,218791808, representing the signal coming from both the target and a neighboring object. The marks and colors are the same as in Figure\,\ref{fig:323174439}. The blue dashed box represents the aperture used in Paper\,I.}
\label{fig:218791808}
\end{figure*}

\begin{figure*}
\includegraphics[width=0.68\textwidth]{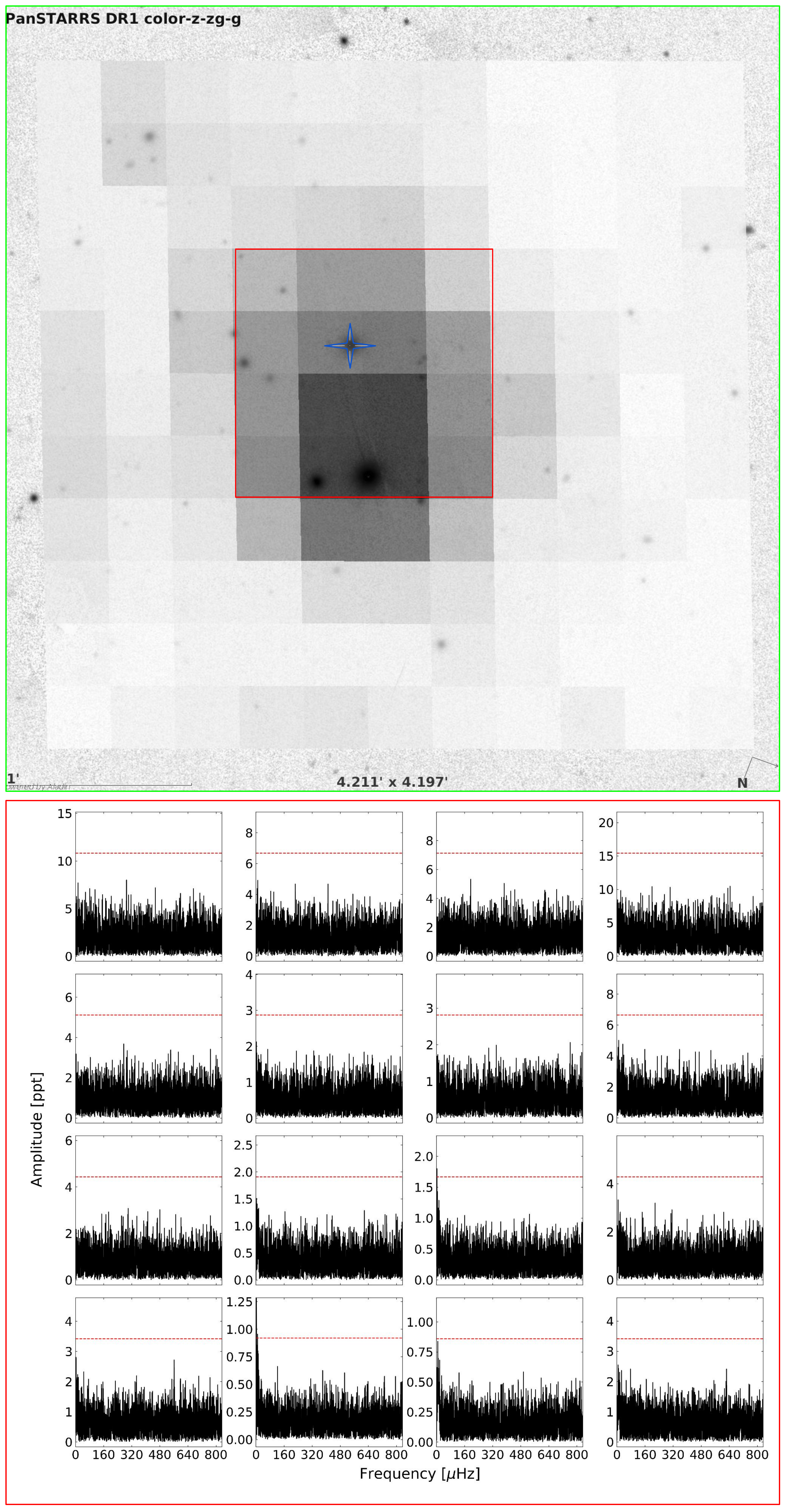}
\caption{Contamination analysis for TIC\,275358553, representing an unconfirmed case. The marks and colors are the same as in Figure\,\ref{fig:323174439} but a PanSTARRS DR1 image is used instead. No significant signal is detected in any of the individual pixels.}
\label{fig:360220395}
\end{figure*}

\begin{figure*}
\includegraphics[width=0.69\textwidth]{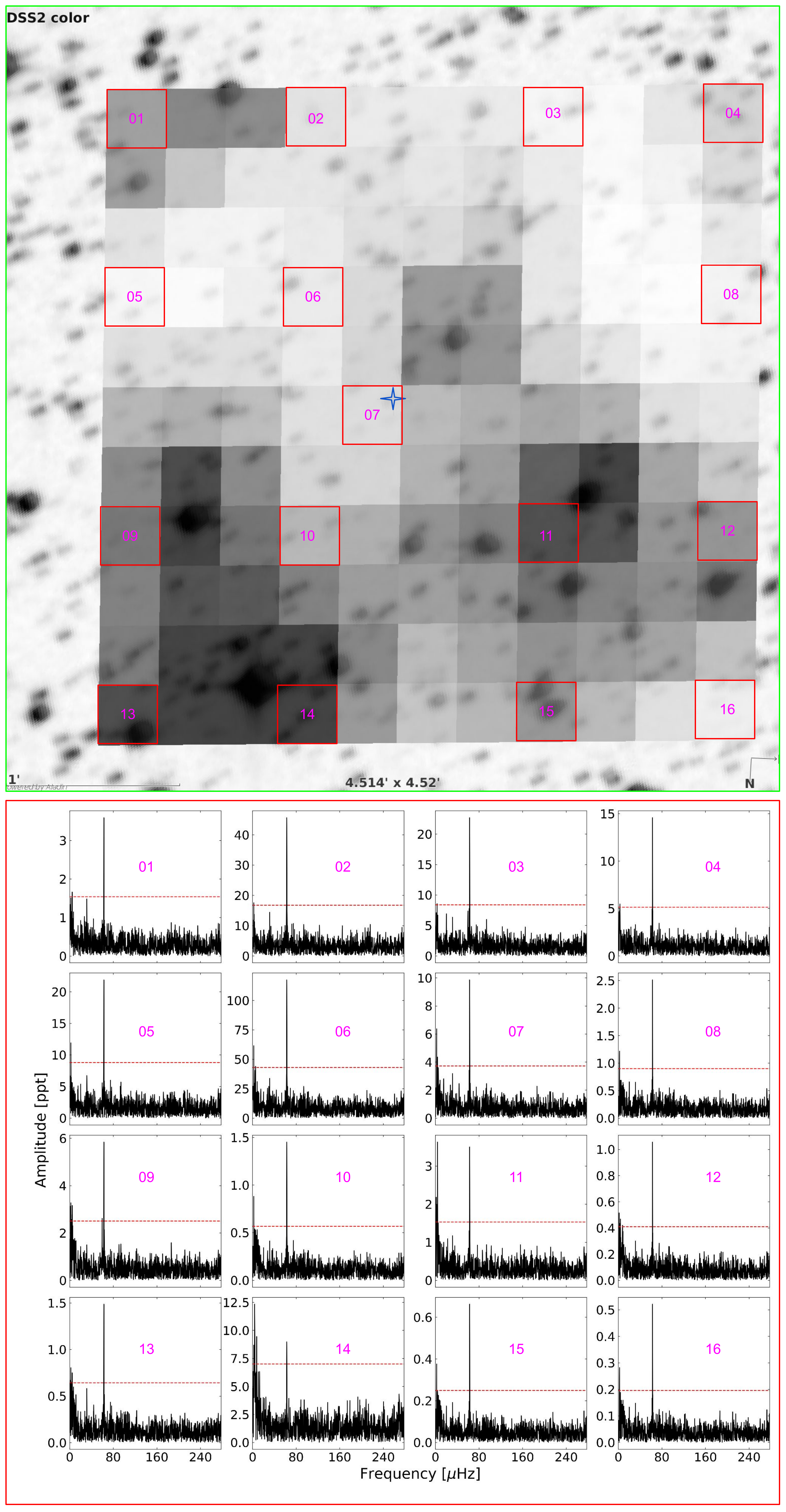}
\caption{Contamination analysis for TIC\,1314011445, representing a false positive case. The marks and colors are the same as in Figure\,\ref{fig:323174439}. Magenta numbers in selected pixels correspond to amplitude spectra shown in the bottom panel. The location of the source of the signal is unconstrained.}
\label{fig:1314011445}
\end{figure*}

\begin{figure*}
\includegraphics[width=0.69\textwidth]{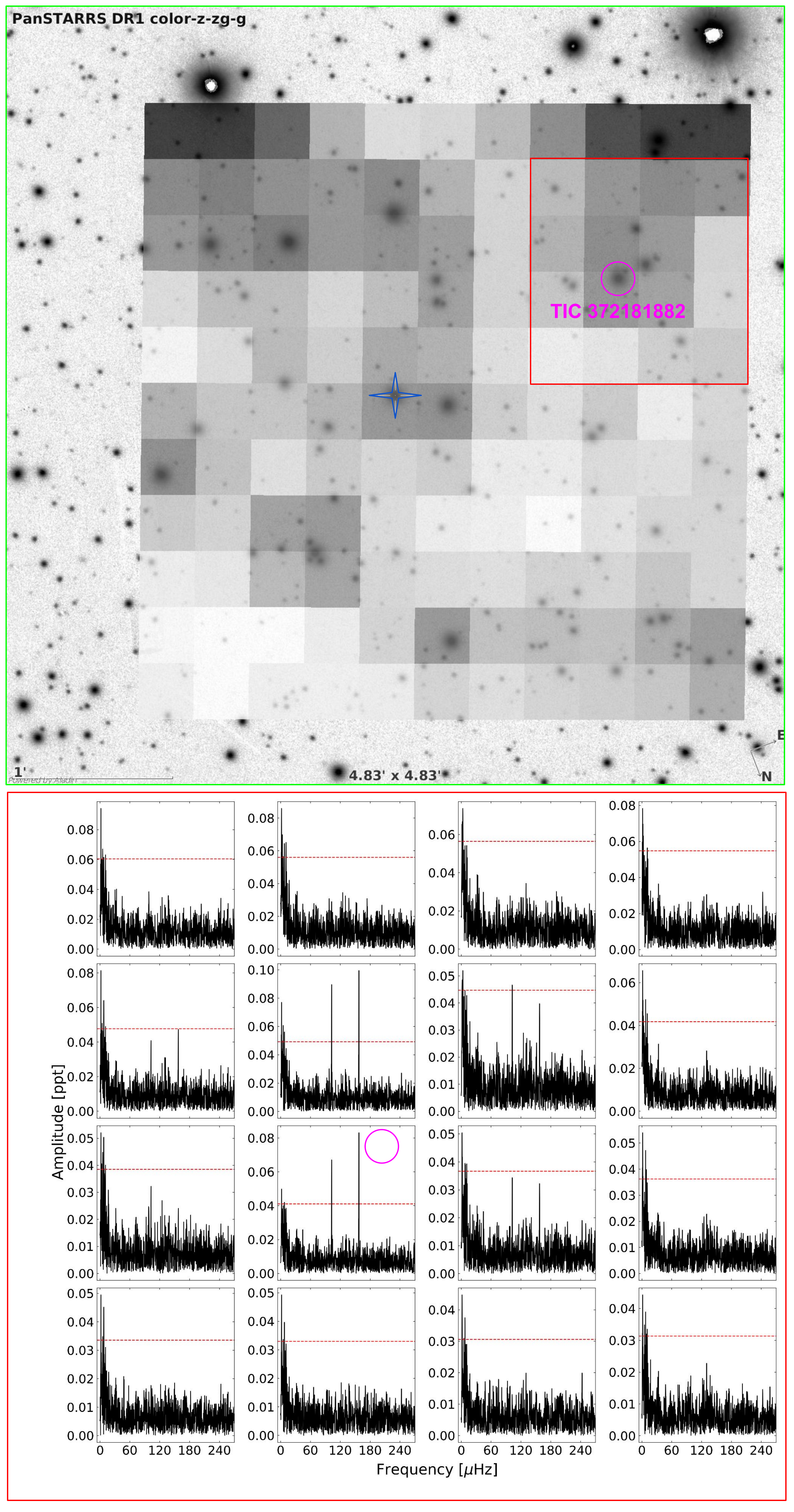}
\caption{Contamination analysis for TIC\,372181885\,(the blue star mark).It is an example of discovering a new non-contaminating variable star (magenta circle) within the target mask during the contamination check of target of interest. The marks and colors are the same as in Figure\,\ref{fig:1514267365}.}
\label{fig:372181885}
\end{figure*}

\begin{itemize}
    \item A signal comes only from the target star, which means no contamination and the target listed in Papers\,I and II is the source of the variability (Figure\,\ref{fig:323174439}). In the \textsc{variable contaminator} column of Tables 1-14, we marked these cases with {\it none}. 

    \item A signal comes from a contaminating object (Figure\,\ref{fig:1514267365}). These cases have the name of the contaminator listed in the \textsc{variable contaminator} column. In parentheses, we added additional information on the contaminators collected from the \textsc{simbad} database. If these contaminators are new variable stars, then they are also listed in Table\,16. This is a false positive case.

    \item A signal comes from both the target and contaminating object(s) (Figure\,\ref{fig:218791808}). These cases will have frequencies assigned to either our target (marked as {\it none}) or contaminator(s) listed in the \textsc{variable contaminator} column. This is the case of a variable target showing an additional signal, which is a false positive.

    \item No signal is detected in single pixels (Figure\,\ref{fig:360220395}). These cases are caused by low S/N in merged pixels defined in Papers\,I and II. In the \textsc{variable contaminator} column we marked these cases with {\it no signal in individual pixels}. This case is not verified, however remarks on nearby stars listed in tables in Papers\,I and II give some clue on the possible contamination.
    
    \item A signal is detected in all pixels across a target mask (Figure\,\ref{fig:1314011445}). In the \textsc{variable contaminator} column we marked these cases with {\it signal in all pixels}. This is a false positive case. The source of the signal is either a nearby bright object that shines over a large area or an instrumental artifact.

    \item A nearby non-contaminating bright object (within the target mask) has been verified positively for its variability (Figure\,\ref{fig:372181885}). These new variable stars are listed in Table\,16.
\end{itemize}

The full tables with remarks of our contamination analysis are presented in the online materials. The most interesting conclusions of our contamination analysis are the following:
\begin{itemize}
    \item Among two sdBV candidates from Paper\,I, TIC\,237597052, is no longer considered a candidate. It turned out to be a main sequence B8 star (Table\,1). Our fit to a spectrum which we collected from the LAMOST survey provides the following atmospheric parameters, \teff\,=\,12\,460(300)\,K, \loggcms\,=\,4.17(12). The other candidate is a confirmed sdB star, so TIC\,262753627 is a new sdB pulsator.
    
    \item Out of 13 sdBV candidates listed in Paper\,II, five are no longer considered the sources of the signal. Apart from these five contaminated targets, TIC\,363766470 is also partially contaminated by ATO\,J265.8117+21.5538. However, our target is still the source of the 116.44\,$\mu$Hz frequency. The contamination of one case could not be verified, while the remaining seven sdBs including TIC\,363766470 are confirmed pulsators. ATO\,J265.8117+21.5538 is listed in the \textsc{simbad} database as an eclipsing binary candidate and we confirm this type by detecting signal at a frequency of 41.09\,$\mu$Hz along with its harmonics. We determined an orbital period of 0.28169\,days (Table\,2). 
    
    \item Among 66 other (than the above) variable sdBs listed in Papers\,I and II, we found 14 not to be the sources of the variable signals. Six cases are not verified. The remaining 46 are confirmed variable sdBs (Tables\,1 and 2).

    \item Out of 10 sdV candidates listed in Paper\,II, four are no longer considered to be the sources of the signals, while TIC\,194781979 was identified as a main sequence B6 star. One case is not verified. The remaining four sdVs are confirmed pulsating candidates (Table\,3).
    
    \item Among 23 other (than the above) variable sdVs listed in Paper\,II, we found seven targets not to be the sources of the signal. Four cases are not verified. The remaining 12 sdVs are confirmed variables (Table\,3).
    
    \item Among 113 spectroscopically unclassified pulsators, we found 68 targets not to be the sources of the signal. Seven cases are not confirmed. The remaining 38 original targets are confirmed pulsators. In the case of TIC\,311792028, the 95.37\,$\mu$Hz  frequency along with its harmonics are native to TIC\,311792021, while the 34.14\,$\mu$Hz frequency originating in our target does not seem to be related to pulsations (Tables\,4 and 5).
    
    \item Among 106 candidate eclipsing binaries, we found 65 targets not to be the sources of the signal. We do not confirm the variability of TIC\,847473488. We found 38 original targets to be confirmed eclipsing binaries. In the case of TIC\,1509561926, the eclipses originate in TYC\,9289-2657-1, while our target itself shows only the 102.89\,$\mu$Hz frequency. In the case of TIC\,159448831, the target is contaminated by an eclipsing binary TIC\,159448824, however our target shows 158.68\,$\mu$Hz along with harmonics. These frequencies, though, are not responsible for the eclipses we plot in Figure\,9 in Paper\,II (Tables\,6 and 7).
    
    \item Among 248 binaries showing one symmetric maximum, we found 118 targets not to be the sources of the signal. Five cases are not verified. We found 125 original targets that are confirmed variables. TIC\,377658867 shows a significant signal at 54.28\,$\mu$Hz and its harmonics, though the contaminator TIC\,378037013 shows 76.27\,$\mu$Hz frequency. Likewise in TIC\,388622589, which shows 13.77\,$\mu$Hz frequency, while a contaminator TIC\,388622573 is responsible for the 161.92\,$\mu$Hz frequency. On the other hand, TIC\,463006021 is heavily contaminated by two sources, even though the target still shows 53.01\,$\mu$Hz frequency and its harmonics. TIC\,463006006 shows the 9.03\,$\mu$Hz frequency presented in Paper\,I, while TIC\,463006054 shows a 6.71\,$\mu$Hz frequency and its harmonics. In the case of TIC\,2040326958, it is actually TIC\,10596964 that shows the 7.87\,$\mu$Hz frequency and its harmonics reported in Paper\,II, though our target still shows a 131.94\,$\mu$Hz frequency (Tables\,8 and 9).
    
    \item Among 52 binaries showing one asymmetric maximum, we found 42 targets not to be the sources of the signal. The remaining 10 original targets are confirmed variables. TIC\,79689537, which shows a 82.64\,$\mu$Hz frequency and its harmonics, is heavily contaminated by two objects. TIC\,79689505 shows the 3.7\,$\mu$Hz frequency and its harmonics reported in Paper\,I, while ATO\,J104.6541-23.0081 shows a 5.56\,$\mu$Hz frequency and its harmonics (Tables\,8 and 9).
    
    \item Among 66 binaries showing two maxima, we found 49 targets not to be the source of the signal. The remaining 17 targets are confirmed binaries (Tables\,8 and 9).
    
    \item The two novae listed in Paper\,I are not contaminated (Table\,10).
    
    \item Among 1\,490 spectroscopically unclassified targets showing signal in their amplitude spectra, we found 719 not to be the sources of the signal while 460 cases are not verified. Such a high number of these cases is a consequence of a low signal that is not detectable in individual pixels. The remaining 311 original targets are confirmed variables. In five cases, our original targets show a signal, although it may not be the one reported in Papers\,I or II (Tables\,11 and 12). 
    
    \item Among 122 spectroscopically classified non-sdB stars, we found 14 targets not to be the sources of the signal. Nine cases are not verified, while the remaining 99 original targets are confirmed variables (Tables\,13 and 14).
\end{itemize}

To summarize our contamination analysis, we found 1\,141 targets not to be the sources of the signal, while 451 targets were not verified. This leaves us with 721 variable sdB candidates remaining, including both pulsating and binary stars.

\section{Updated amplitude spectra and new pulsating sdB stars}
\raggedbottom
We took advantage of access to the 10\,min FFI data collected during Years\,3 and 4 to discover new variable sdB stars and to improve the amplitude spectra for known sdBV stars. Not all variable stars listed in Papers\,I and II were subject to this updated analysis. Binaries would surely benefit from a three times better sampling, which would yield much better definition of eclipses and more precise orbital period estimation, but it would not affect their variability type. The contaminated stars, which turned out not to be sources of the signal, were also excluded. We focused only on targets that show rich signal close to the 277.7\,\uHz\ Nyquist frequency, which appear to be quite convincing pulsations. In these cases shifting the Nyquist frequency to a three times higher value would uncover the entire g-mode region of possible pulsating sdB stars. We ended up with a final list of 78 targets.

\begin{figure*}
\includegraphics[width=0.83\textwidth]{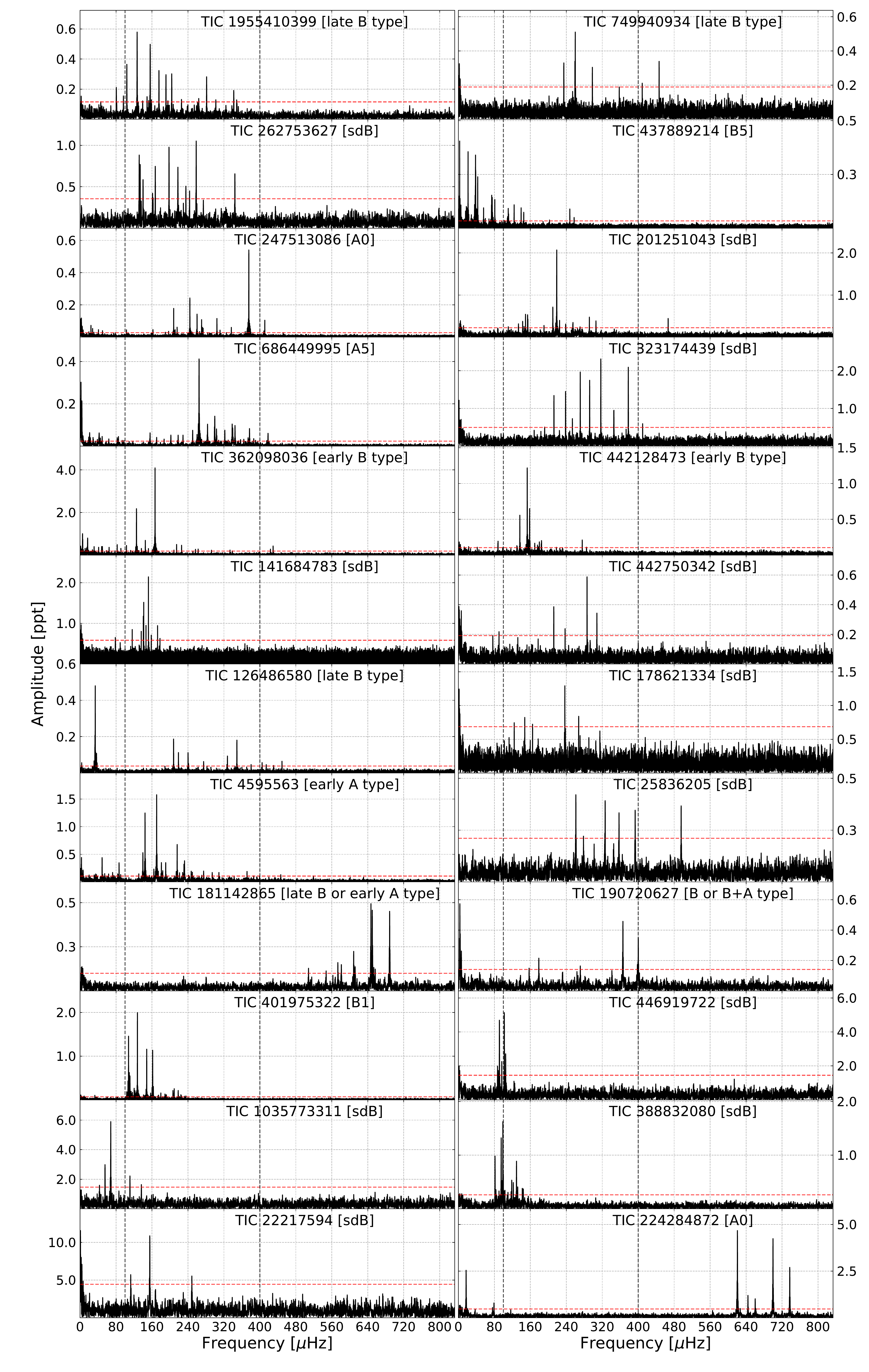}
\caption{The amplitude spectra of pulsators analyzed with 10\,min FFI data. Vertical dashed lines indicate the typical g-mode region of pulsating hot subdwarfs. The horizontal lines mark the detection thresholds.}
\label{fig:pulsators}
\end{figure*}

The light curve extraction process and data reduction of the 10\,min FFI data, are the same as for the 30\,min FFI data described in Papers\,I and II, where we refer the reader for details. Out of 78 pre-selected targets we found 24 that show multiple frequencies, which we interpret as pulsations, and the number of frequencies were significantly increased or more frequencies beyond the 277.7\,\uHz\ Nyquist frequency were detected. We list these targets in Table\,15 in the online material. The targets are confirmed as the original sources of the detected signals. We show the amplitude spectra of these 24 targets in Figure\,\ref{fig:pulsators}. We marked the typical g-mode frequency range ({\it i.e.} 100\,--\,400\,\uHz) with dashed vertical lines. This region overlaps with the typical p-mode frequency range of $\delta$\,Scuti stars, so it is not generally straightforward to claim the detection of a pulsating sdB star based only on the amplitude spectrum content. Alternatively, if the frequencies detected are outside the indicated range, we might well doubt the detection of an sdB pulsator. For instance, TIC\,224284872 is an A star and the frequencies are outside the expected range. Similarly, TIC\,181142865 turned out to be a main sequence star. TIC\,437889214 shows frequencies below the expected region -- its spectral type is B5. There are other targets which are not classified as hot subdwarfs, yet show frequencies in the range characteristic of pulsating sdB stars. These cases may be either $\delta$\,Scuti pulsators or misclassified hot subdwarfs. We confirm a detection of g-mode pulsations in 11 sdB pulsators. TIC\,442750342 seems to be an exception in our sample being a low gravity and cool sdB pulsator.

\begin{table*}
\centering
\caption{Pulsators confirmed with the 10\,min FFI data. We show amplitude spectra of these objects in Figure\,\ref{fig:pulsators}. This table is also included in the on-line materials.}
\label{tab:tab15}
\rowcolors{1}{}{light3}
{\begin{tabular}{clrccc}
\hline
\rowcolor{light1}
& & & \multicolumn{1}{c}{G} & \multicolumn{1}{c}{Variable} & \\
\rowcolor{light1}
\multicolumn{1}{c}{\multirow{-2}{*}{No.}} & \multicolumn{1}{c}{\multirow{-2}{*}{\gaia\ DR2}} & \multicolumn{1}{c}{\multirow{-2}{*}{TIC}} & \multicolumn{1}{c}{[mag]} & \multicolumn{1}{c}{contaminator} & \multicolumn{1}{c}{\multirow{-2}{*}{Remarks}}\\
\hline
1 & 1861191062326013696 & 1955410399 & 10.66 & none & late B type \\
2 & 3032890473180888192 & 749940934 & 12.46 & none & late B type \\
3 & 3159937564294110080 & 262753627 & 12.46 & none & sdB \\
4 & 3344114626761364224 & 437889214 & 10.16 & none & B5 \\
5 & 3396397877830881792 & 247513086 & 8.44 & none & A0 \\
6 & 4923853724788504192 & 201251043 & 11.93 & none & sdB \\
7 & 5090382015016433920 & 686449995 & 6.99 & none & A5 \\
8 & 5196271513123121152 & 323174439 & 13.30 & none & sdB \\
9 & 5250674622612902912 & 362098036 & 11.48 & none & early B type \\
10 & 5257747299878049152 & 442128473 & 10.84 & none & early B type \\
11 & 5266133451162548864 & 141684783 & 14.53 & none & sdB \\
12 & 5307946881949072000 & 442750342 & 12.82 & none & sdB \\
13 & 5326745919424790656 & 126486580 & 9.30 & none & late B type \\
14 & 5362558250096941056 & 178621334 & 13.33 & none & sdB \\
15 & 5429254969036524416 & 4595563 & 10.11 & none & early A type \\
16 & 5439887654492064256 & 25836205 & 13.14 & none & sdB \\
17 & 5525342630213336448 & 181142865 & 11.11 & none & late B or early A type \\
18 & 5622554881336253824 & 190720627 & 11.07 & none & B or B+A type \\
19 & 5796399012705196800 & 401975322 & 10.21 & none & B1 \\
20 & 5823403087017696384 & 446919722 & 12.96 & none & sdB \\
21 & 5872410931625754624 & 1035773311 & 17.46 & none & sdB \\
22 & 5922070855307705472 & 388832080 & 12.58 & none & sdB \\
23 & 6143764182206682112 & 22217594 & 15.16 & none & sdB \\
24 & 6534581776366266752 & 224284872 & 13.53 & none & A0 \\
\hline
\end{tabular}}
\end{table*}

\section{New variables}                         
As a by-product of the contamination analysis we report the true sources of variability preliminarily assigned to the targets listed in Papers\,I and II. In Tables\,1-14 we provide a \textsc{variable contaminator} column which, in the case of a positive variability contamination, contains a name of a contaminator. In addition, we detected new variables that do not contaminate our pre-selected targets but are located within the target masks of our targets. For a practical reason, all these variable contaminators and new non-contaminating variables are listed in Table\,16 in the online materials. In total, we report detection of 682 new variable stars, including two, listed last in Table\,16, that have no \gaia\ \citep{gaia18} designation yet. To be precise, the discovery of the variability of the majority of these stars was presented in Papers\,I and II, so only 97 stars are found to be new variables (accounting for Papers\,I and II) while the remaining 585 stars now have the variability properly assigned (as compared to Paper\,I and II).

\begin{table*}
\centering
\caption{Basic information of the 682 new variable stars found during the contamination analysis in both SEH and NEH. Only first five objects are listed while the full table can be found in the online materials.}
\label{tab:tab16}
\rowcolors{1}{}{light3}
{\begin{tabular}{clrlc}
\hline
\rowcolor{light1}
& & & & \multicolumn{1}{c}{G} \\
\rowcolor{light1}
\multicolumn{1}{c}{\multirow{-2}{*}{No.}} & \multicolumn{1}{c}{\multirow{-2}{*}{\gaia\ DR2}} & \multicolumn{1}{c}{\multirow{-2}{*}{TIC}} & \multicolumn{1}{c}{\multirow{-2}{*}{Name}} & \multicolumn{1}{c}{[mag]}\\
\hline
1 & 1000847845211000960 & 14196021 & - & 16.65 \\
2 & 1030011910101662336 & 467154863 & - & 12.56 \\
3 & 1082306439760980224 & 284473271 & - & 16.95 \\
4 & 1113516077316307328 & 468921975 & - & 17.32 \\
5 & 1131845245388039296 & 459183003 & - & 14.38 \\
\hline
\end{tabular}}
\end{table*}

\section{Summary}
We presented the results of our contamination analysis of stars included in Papers\,I and II. We identified 1\,141 false positives, while 451 variables were not verified because, in most cases, the signal is of too low amplitude to be detected in individual pixels. The total number of targets, which are the sources of the signal we presented in Papers\,I and II is 721. As a by-product of our contamination analysis we found 97 new variables that happened to be within target masks of our original stars listed in Papers\,I and II. In total, we analysed 2\,995 targets in TESS fields, where 2313 targets were presented in Papers\,I and II and the remaining 682 variable targets were found during the contamination check. Out of the 2\,313 targets, we confirmed 721 as variable after contamination analysis. Hence the total number of variable targets we found is 1\,403.

We pre-selected 78 uncontaminated targets that are pulsator candidates (that is, they show rich pulsation content close to the 277.7\,\uHz\ Nyquist frequency) for additional analysis using the 10\,min FFI data collected during Years\,3 and 4. We ended up with 24 targets for which those new data turned out to be beneficial -- that is, more peaks either below, or especially beyond, the 277.7\,\uHz\ frequency were detected. For any of these 24 stars without spectral type, we used publicly available data and/or spectroscopic data collected with the 1.9\,m telescope at the South African Astronomical Observatory to identify hot subdwarfs. In total, we found 11 new sdB pulsators. Details of the spectroscopic analysis will be provided in Worters et al. (in preparation).

One of the pulsator candidates, TIC\,362098036, was a subject of a pulsation mode identification and the result was reported in Paper\,1. Our analysis confirmed that the target is a main sequence B star, which makes the mode identification irrelevant.

\section*{Acknowledgements}
Financial support from the National Science Center in Poland under projects No.\,UMO-2017/26/E/ST9/00703 and UMO-2017/25/B/ST9/02218 is acknowledged. 
PN acknowledges support from the Grant Agency of the Czech Republic (GA\v{C}R 22-34467S). The Astronomical Institute in Ond\v{r}ejov is supported by the project RVO:67985815. This paper includes data collected by the \tess mission. Funding for the \tess\ mission is provided by the NASA Explorer Program. This work has made use of data from the European Space Agency (ESA) mission \gaia\ (\url{https://www.cosmos.esa.int/gaia}), processed by the \gaia\ Data Processing and Analysis Consortium (DPAC, \url{https://www.cosmos.esa.int/web/gaia/dpac/consortium}). Funding for the DPAC has been provided by national institutions, in particular, the institutions participating in the \gaia\ multilateral agreement.
This research has used the services of \url{www.Astroserver.org}.

\section*{Data availability}
The datasets were derived from MAST in the public domain archive.stsci.edu.
%%%%%%%%%%%%%%%%%%%%%%%%%%%%%%%%%%%%%%%%%%%%%%%%%%



\bibliographystyle{mnras}
\bibliography{bibliography.bib} % if your bibtex file is called example.bib

%\printbibliography
%\begin{thebibliography}{99}
%\bibitem[\protect\citepauthoryear{Author}{2012}]{Author2012}
%\bibitem[\protect\citepauthoryear{Others}{2013}]{Others2013}
%\end{thebibliography}

%%%%%%%%%%%%%%%%%%%%%%%%%%%%%%%%%%%%%%%%%%%%%%%%%%

%%%%%%%%%%%%%%%%% APPENDICES %%%%%%%%%%%%%%%%%%%%%

%\appendix



%%%%%%%%%%%%%%%%%%%%%%%%%%%%%%%%%%%%%%%%%%%%%%%%%%

% Don't change these lines
\bsp	% typesetting comment
\label{lastpage}
\end{document}